\documentclass[conference]{IEEEtran}
\IEEEoverridecommandlockouts
\usepackage{cite}
\usepackage{amsmath,amssymb,amsfonts}
\usepackage{amsmath}
\usepackage{algorithm}
\usepackage{algorithmic}
\usepackage{graphicx}
\usepackage{textcomp}
\usepackage{bm} 
\usepackage{xcolor}
\usepackage{booktabs}
\usepackage{subfig}
\def\BibTeX{{\rm B\kern-.05em{\sc i\kern-.025em b}\kern-.08em
T\kern-.1667em\lower.7ex\hbox{E}\kern-.125emX}}
\begin{document}

\title{Simulation of Emergency Evacuation in Large Scale Metropolitan Railway Systems for Urban Resilience}

\author{
    \IEEEauthorblockN{
        Hangli Ge\IEEEauthorrefmark{1}, Xiaojie Yang\IEEEauthorrefmark{2}, Zipei Fan\IEEEauthorrefmark{3}, 
  Francesco Flammini\IEEEauthorrefmark{4}, Noboru Koshizuka\IEEEauthorrefmark{1}
    }
    \IEEEauthorblockA{
        \IEEEauthorrefmark{1}Interfaculty Initiative in Information Studies, The University of Tokyo, Tokyo, Japan\\
        \IEEEauthorrefmark{2}Graduate School of Interdisciplinary Information Studies, The University of Tokyo, Tokyo, Japan\\
         \IEEEauthorrefmark{3}School of Artificial Intelligence, Jilin University, Jilin, China\\
        \IEEEauthorrefmark{4}Department of Mathematics and Computer Science Ulisse Dini, University of Florence, Florence, Italy,\\
        and IDSIA USI-SUPSI, University of Applied Sciences and Arts of Southern Switzerland, Lugano, Switzerland\\
        Email: \{hanglige, xiaojieyang, qkoshi\}@g.ecc.u-tokyo.ac.jp,\\
        fanzipei@jlu.edu.cn, francesco.flammini@unifi.it
    }
}

\maketitle

\begin{abstract}
This paper presents a simulation for traffic evacuation during railway disruptions to enhance urban resilience. The research focuses on large-scale railway networks and provides flexible simulation settings to accommodate multiple node or line failures. The evacuation optimization model is mathematically formulated using matrix computation and nonlinear programming. The simulation integrates railway lines operated by various companies, along with external geographical features of the network. Furthermore, to address computational complexity in large-scale graph networks, a subgraph partitioning solution is employed for computation acceleration. The model is evaluated using the extensive railway network of Greater Tokyo. Data collection included both railway network structure and real-world GPS footfall data to estimate the number of station-area visitors for simulation input and evaluation purposes. Several evacuation scenarios were simulated for major stations including Tokyo, Shinjuku,  Shibuya and so on. The results demonstrate that both evacuation passenger flow (EPF) and average travel time (ATT) during emergencies were successfully optimized, while remaining within the capacity constraints of neighboring stations and the targeted disruption recovery times.

 
\end{abstract}





\section{Introduction}
Railway systems, as critical transportation infrastructures, significantly contribute to economic growth and urban development. However, as the railway network grows more complex and passenger volumes rise, it becomes more vulnerable to disruptions. Under abnormal or emergency conditions, such as natural disasters, equipment breakdown, cyber-attacks \cite{bellini2025situation}, operational failures, accidents, etc, trains may fail to run according to schedule or may even face disruptions, resulting in significant variations in service, including congestion, stoppages, or even accidents.

Improving the resilience of railways to recover to normal conditions more quickly is a critical problem in the management of railway systems \cite{chen2024resilience,esmalian2022operationalizing}. Among these recovery practices, traffic analysis and evacuation simulation are significant approaches \cite{ge2025trafficai4rails}. They can inform evacuation strategies and policies to improve safety and efficiency. \textbf{However, these tasks are highly challenging due to the lack of readily available data on railway networks and passenger flows on a large scale (cross-city, nation wide or even cross-country)}. If the model focuses solely on a single railway line or lines operated by the same company, the process is relatively straightforward. However, it becomes particularly complex when we need to construct the entire rail network at large scale. The federation of multiple railway operators further complicates data collection, integration, and the simulation of decision-making within a comprehensive and large-scale railway network.


\begin{figure}[t]
\centering
\includegraphics[height=2.3in, width=0.7\linewidth]{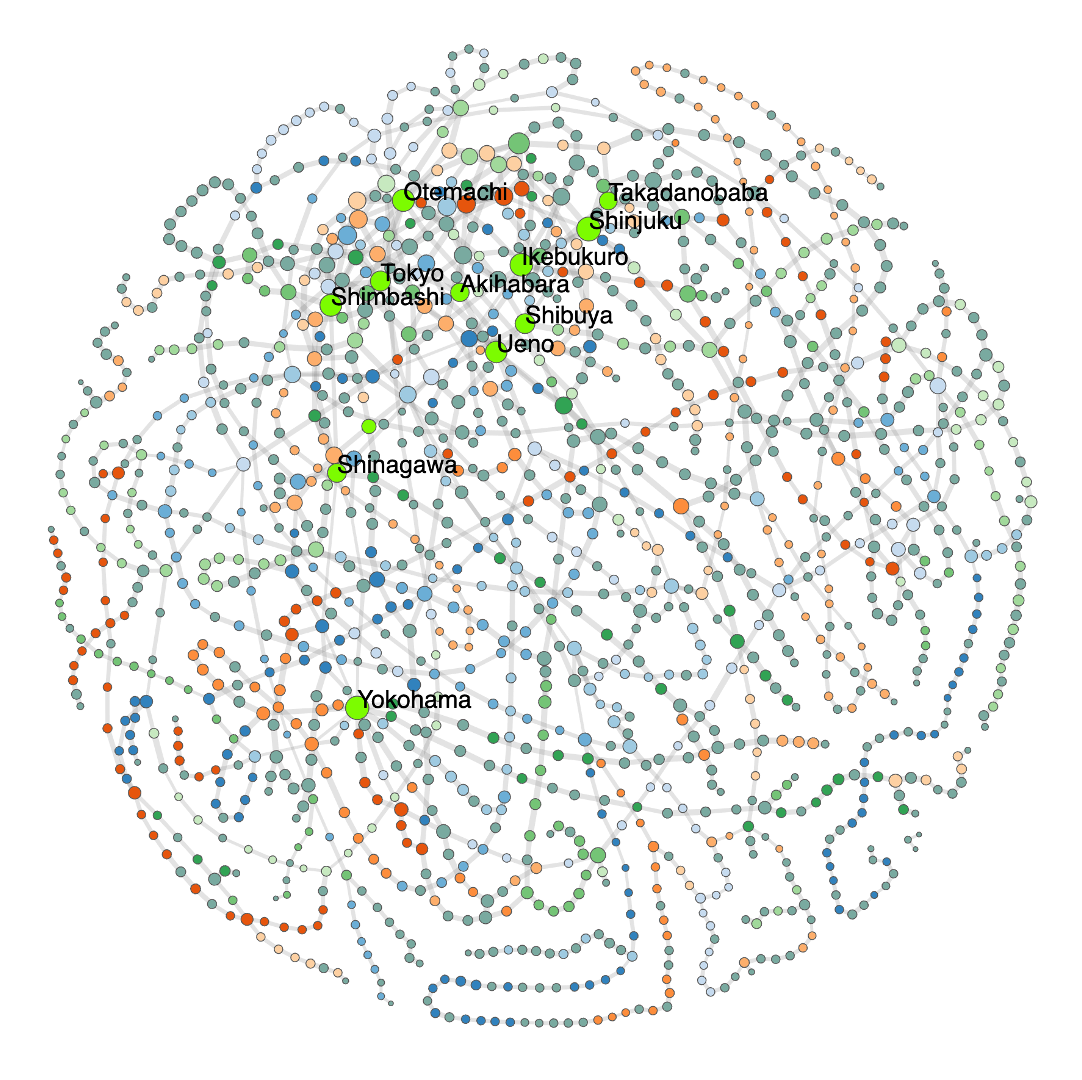}
\caption{The large scale of the Greater Tokyo railway system}
\label{fig:net}
\end{figure}


In this paper, we propose a traffic evacuation simulation approach for large scale railway networks. It focuses on the failures of several nodes or lines in the network topology. \textbf{Our proposed model optimizes traffic evacuation considering delayed time, overcrowded passenger flows, and the capacities of neighboring stations}. We collected large-scale railway network data from the Greater Tokyo area (as the visualization shown in Figure \ref{fig:net}), incorporating data from different operating companies and geographical features such as connectivity, distance, which were modeled in the evacuation cost matrix. The traffic evacuation model includes an optimization algorithm based on matrix calculations, making it both intuitive and interpretable. Therefore, the contributions of this work can be summarized as follows:


\begin{itemize}
    \item To fully exploit useful features, we  proposed an evacuation cost model of multi-features fusion, which also dynamically changes based on predefined disruption time. The optimization model considers the capacity of surrounding stations and evacuation cost. It is able to flexibly simulate disruptions at multiple stations or lines, enhancing its applicability. 
    \item The model uses matrix-based calculations and optimization, making it intuitive, straightforward and simple to deploy. We also employed a graph partition strategy to accelerate the computation of large scale graph.
   \item We conducted case studies on Greater Tokyo’s large-scale railway network for validating the effectiveness. The geographic features and passenger statistics are utilized for simulation on several major stations. 
\end{itemize}




\section{Related Work}


Overall, traffic evacuation simulation is a powerful tool for evaluating and optimizing evacuation plans, and it can be used to simulate various aspects of the evacuation process, including driver behavior, traffic flow, and road network conditions. By using traffic simulation models and evacuation planning algorithms, traffic evacuation simulation can help emergency responders and planners develop effective evacuation plans that minimize congestion and ensure the safety of evacuees. One such model is proposed in \cite{ge2025trafficai4rails}, which focuses on optimizing evacuation strategies in response to disruptions affecting multiple nodes within the rail network.

Traffic evacuation models have also been proposed for optimizing the evacuation process, minimize travel delays, and ensure the safety of passengers. There are several types of traffic evacuation models, including microscopic, macroscopic, and mesoscopic models \cite{intini2019traffic}. Several practices are used to improve the resilience of railway systems. They put the focus on transport planning mainly include the identification of critical nodes and edges \cite{lee2022quantitative}, transport network design \cite{xu2021enhancing}, emergency response, and evacuation \cite{esmalian2022operationalizing}. A study proposed a novel method to optimize the resilience of urban rail systems, focusing on the combined impact of delayed trains and overcrowded passenger flows \cite{li2023optimization}.
A resilience-based optimization model, which combines resilience evaluation and dispatching optimization was proposed to maximize the resilience index of evacuated passengers of bus system \cite{Zhang2023}. A similar approach is to represent an integrated network, generate candidate bus-bridging routes using the K-shortest paths algorithm, and solve the optimization model to determine the optimal train allocation among the candidate routes \cite{Jaber2024}. The proposal was applied to a case study in the Ile de France region, Paris and suburbs, to optimize transportation management during interruptions.

In summary, data-driven approaches for simulating traffic evacuation are significant to improve the resilience of railway systems. However, it requires a holistic view that considers various components, including infrastructure, external features of users or cities, operator organization, integration level, for emergency risk management and recovery. The Tokyo railway network with large scale is a complex and dynamic system that requires careful planning and analysis to optimize passenger flow and reduce congestion \cite{Zhang2021}. This needs to be achieved through the development of integrated data management systems that bring together various stakeholders, including operators, maintenance personnel such as local government, and IT-service providers and so on. More features of the rail network are modeled, the more constraints of the optimization solutions align with real-world conditions, makes the 
optimization solutions more feasible.

\section{Problem Definition and Notations}

The railway traffic evacuation problem was mathematically defined and optimized using nonlinear programming. 

\subsection{Input of Railway Network}
\begin{table}[h]
    \centering
    \renewcommand{\arraystretch}{1.5} 
    \begin{tabular}{|c|c|l|}
        \hline
        \textbf{Symbol} & \textbf{Size} & \textbf{Description} \\
        \hline
        \( \bm{Y} \) & \( \mathbb{R}^{n \times 1} \) & Passenger count at each station. \\
        \hline
        \( \bm{J}\) & \( \mathbb{R}^{n \times 1} \) & Station status (1: blocked, 0: normal). \\
        \hline
       \( \bm{\bar{J}} \) & \( \mathbb{R}^{n \times 1} \) & Reverse of $\bm{J}$ \\
        \hline
        \( \bm{X} \) & \( \mathbb{R}^{n \times 1} \) & Increased capacity of station volume \\
        \hline
        \( \bm{A_{\text{con}}}, \bm{A_{\text{dis}}}, \bm{A_{\text{cost}}} \) & \( \mathbb{R}^{n \times n} \) & Connectivity, distance, and cost matrices. \\
        \hline
        \( T_{lm} \) & \( 1\) & Disrupted time (minutes), bigger than 0 \\
        \hline
        \( \bm{T_{\text{train}}}\) & \( \mathbb{R}^{n \times n} \) & Train moving time among the station nodes\\
        \hline
    \end{tabular}
    \caption{Summary of Notations}
\end{table}

Suppose we have \( n \) stations, represented by the set of station nodes \( \{ v_{1}, v_{2}, ..., v_{n-1}, v_{n} \} \). Hereafter, we define the following notations:


\textbf{Connectivity Matrix of $\bm{A_{con} \in \mathbb{R}^{n \times n}}$ }: Connectivity is the most basic principle to ensure the completeness and correctness of a railway network. Therefore, each element $a_{i,j}$ in $\bm{A_{con}}$ is denoted by Equation~\ref{eq:w_c}.
\begin{equation}
a_{i,j} =
\label{eq:w_c} 
\begin{cases}
     1   &   \text{if $v_{i}$ connects to $v_{j}$ in the railway line}\\  
     0  & otherwise\\
\end{cases}
\end{equation}



\textbf{Distance Matrix of $\bm{A_{dis}} \in \mathbb{R}^{n \times n}$}: Let distance matrix be $\bm{A_{dis}}$, and each element $a_{i,j}$ is the Haversine distance between two points (marked with latitude and longitude) on the globe.



\subsection{Dynamic Cost Matrix Generation}

\begin{figure}[t]
\centering
\includegraphics[scale=0.4]{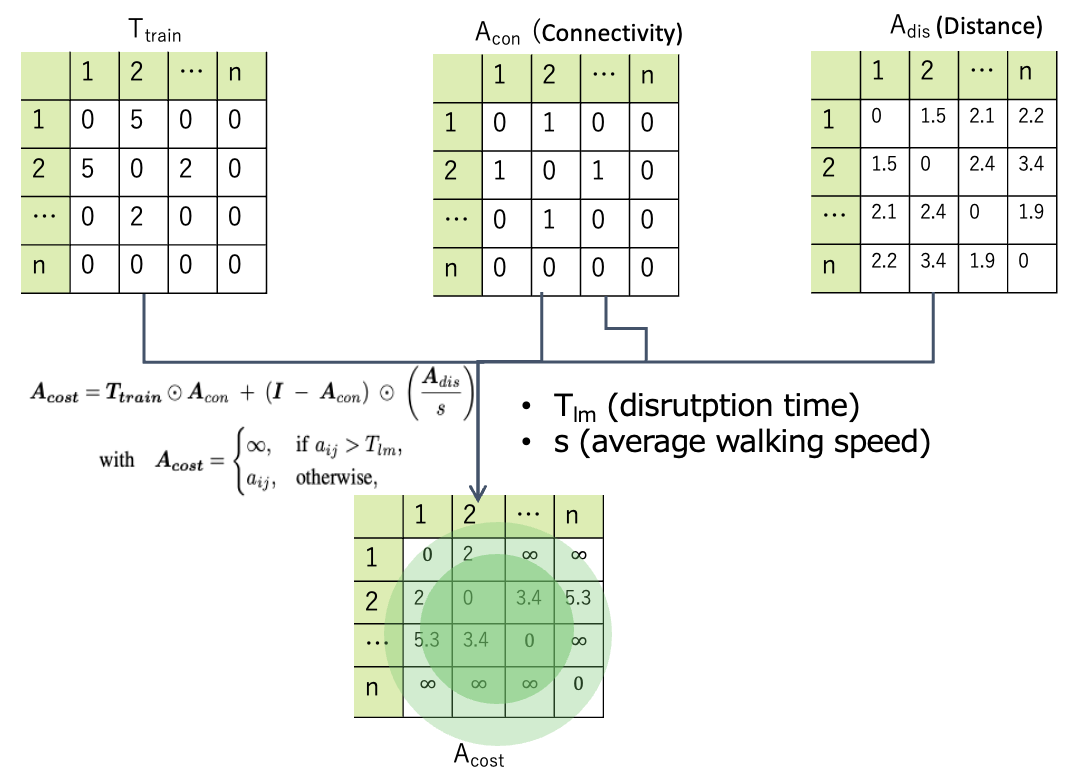}
\caption{Dynamic cost matrix generation based on fusion of connectivity and distance matrix}
\label{fig:dycost}
\end{figure}

The travel cost matrix was constructed based on two types of basic matrix: connectivity matrix and distance matrix. To fully exploit different but useful heterogeneous spatial correlation among stations, the fused cost matrix (denoted as $A_{cost}$) were calculated by fusing the $A_{con}$  and $A_{dis}$ matrices at the element level, as shown in Equation~\ref{eq:a_cost}. It is also illustrated in Figure~\ref{fig:dycost}.
\begin{equation}
\begin{aligned}
    \bm{A_{cost}} &= \bm{T_{train}} \odot \bm{A}_{con} \;+\; (\bm{I} \;-\; \bm{A}_{con}) \;\odot\; \left(\frac{\bm{A}_{dis}}{s}\right) \\
    &\quad \text{with} \quad
    \bm{A_{cost}} = \begin{cases} 
      \infty, & \text{if } a_{ij} > T_{lm}, \\
      a_{ij}, & \text{otherwise},
    \end{cases}
\end{aligned}
\label{eq:a_cost}
\end{equation}

Throughout the formulation, the operator ``$\odot$'' denotes the Hadamard (element-wise) product. where $\bm{I}$ is the identity matrix. The matrix $\bm{T_{train}}$ represents the estimated travel cost between any pair of stations within the rail transit network. $\bm{T_{train}} \in \mathbb{R}^{n \times n}$ where each element $t_{i,j}$ denotes the shortest travel time by train from station $v_{i}$ to $v_{j}$. To determine the travel cost, we calculate the minimum travel time between all pairs of stations with one transfer (either walking or taking a train). $\left(\frac{\bm{A}_{dis}}{s}\right)$ denotes the walking time between two stations and we set the average walking speed is $s$ (set to be 5km~/h). In particular, during our optimization, we consider it impractical for evacuation time to exceed the station's recovery time. Therefore, we set the entry $a_{i,j}$ in $\bm{A_{cost}}$ that exceeds a certain disruption $T_{lm}$ (e.g., 30 minutes or 60 minutes) to infinity, thus helping us quickly narrow down to reasonable candidate stations during the computation.

Algorithm~\ref{alg:train} shows the generation of matrix $\bm{T_{train}}$. We first construct the rail network as a graph using the \texttt{NetworkX} library, where each node represents a station and contains attributes such as geographic coordinates and line affiliation. The distance between adjacent nodes is computed using the haversine formula, and the in-train travel time is estimated with the expression $x /v$, where $x$ denotes the distance in meters and $v$ refers to the speed of the train ($800 \ meters/minute$). In addition, we assume a fixed stop time of $t_s$ minute at each station (1 minute here). For any pair of stations, we compute the shortest path and accumulate the total travel time, including both in-train time and stop time. If a line transfer occurs along the path, an additional change time of $t_c$ minutes (7.5 minutes here) is added to the travel cost to reflect the transfer penalty. This approach results in a comprehensive travel cost matrix that accounts for spatial distance, operational speed, and transfer delays within the network.

\begin{algorithm}
\caption{Generate Train Travel Cost Matrix $T_{train}$}
\label{alg:train}
\begin{algorithmic}

\REQUIRE Connection matrix $A_{con}$, coordinates $\text{Coord}$, line labels $\text{LineLabel}$
\ENSURE Travel cost matrix $T_{train}$

\STATE Initialize graph $G$ with nodes from $A_{con}$
\FOR{each station $i$}
    \STATE Assign $\text{Coord}[i]$ and $\text{LineLabel}[i]$ to node $i$
\ENDFOR

\FOR{each connected pair $(i, j)$ in $A_{con}$}
    \STATE Compute distance $d_{ij}$ using haversine formula
    \STATE Compute time $t_{ij} \leftarrow d_{ij} / v$
    \STATE Add edge $(i, j)$ to $G$ with weight $t_{ij}$
\ENDFOR

\STATE Initialize $T_{train}$ as a zero matrix
\FOR{each station pair $(s, t)$}
    \STATE Find shortest path $P$ from $s$ to $t$ in $G$
    \STATE $stops \leftarrow |P| - 1$ 
    \STATE $changes \leftarrow |LineLabel \in P| - 1$
    \STATE $C \leftarrow t_{s} \times stops + t_{c} \times changes + \sum_{i, j \in P}t_{ij}$ 
    \STATE Set $T_{train}[s][t] \leftarrow C$
\ENDFOR

\end{algorithmic}
\end{algorithm}

\subsection{Objective and Output}
Let the matrix $\bm{K} \in \mathbb{R}^{n \times n}$ be the optimized traffic evacuation matrix after the shutdown of several stations, where $k_{i,j}$ denotes the count of passengers need to be evacuated from the $i$-th station to the $j$-th station. The objective is to find a traffic assignment output \( \textbf{K} = \{k_{ij}\} \) to evacuate the flow of passengers from affected stations. 
The solution must ensure efficient passenger evacuation while preserving network stability and minimizing evacuation cost. The problem was formulated that simultaneously minimizes the average evacuation cost subject to flow conservation and capacity constraints. Therefore, the evacuation strategy must satisfy the following constraints and optimization objectives.
\subsection{Constraints}
\noindent \textbf{(1) Equality Constraint:} The total number of rerouted passengers must match the affected passenger volume:  
\begin{equation}
    \bm{J} \odot \bm{Y} 
\;=\;
\bm{K}\bm{J}.
\end{equation}
This matrix computation ensures an equivalent relation between $\bm{K}$ and the given data, which means that all people in blocked stations must leave for other stations.

\noindent \textbf{(2) Inequality Constraint:}
\textbf{Capacity Constraint:} All the stations have a common constant ratio for upper limit capacity:  
\begin{equation}
  \bm{Y} \odot \bm{\bar{J}}
\;+\;
\bm{K}^{\top} \odot \bm{\bar{J}}
\;\;\le\;\;
\bm{X} \,\odot\, \bm{Y} \,\odot\, \bm{\bar{J}}.
\end{equation}
This can be interpreted as a limit on the volume in each station, meaning that each blocked station has a fixed capacity $\bm{X}$ for a certain number of people from other stations.

\noindent \textbf{(3) Diagonal Constraint:}
\begin{equation}
    k_{i,i} = 0,
\quad
\forall \, i = 1,\dots,n.
\end{equation}
Each diagonal entry $k_{i,i}$ of $\bm{K}$ must be zero, which means that a station cannot evacuate to itself. It prevents meaningless self-flows in the matrix and ensures that all movements are toward alternative stations.

\noindent \textbf{(4) Additional Row/Column Constraints Depending on \(\bm{J}\):} This rule provides a more detailed constraint based on prior knowledge. For each \(i \neq j\), the values of \(k_{i,j}\) and \(k_{j,i}\) must respect certain bounds depending on whether \(J_{i}\) is 0 or 1. In particular:
\begin{equation}
\begin{cases}
k_{*,i} \;\ge\; 0, \quad k_{i,*} = 0 &\quad  \text{if $J_{i} = 0$},\\
k_{i,*} \;\ge\; 0, \quad k_{*,i} = 0 &\quad  \text{if $J_{i}= 1$}.\\
\end{cases}
\end{equation}
For station $i$ in the normal status: inflow is possible; outflow is not allowed.
otherwise, station $i$ is blocked, outflow is mandatory and inflow is not allowed.

\subsection{Optimization}

\textbf{Travel Cost Minimization:} To minimize the travel time or walking distance caused by evacuation, we have $\bm{A_{cost}}$ representing the travel cost matrix within all stations (each element denoted as  \( a_{ij} \)). The optimization seeks to minimize the average travel cost for each passenger. The objective function is the mean value of the element-wise product between $\bm{A_{cost}}$ and $\bm{K}$ (denoted as $(\bm{A_{cost}} \odot \bm{K})$), which represents the average travel cost of each person according to the decision matrix $\bm{K}$. We then multiply this by $\bm{J}$ and $\bm{J}^{\top}$ to obtain a scalar:

\begin{equation}
sum_{AK} = \sum_{i}(\bm{A_{cost}} \odot \bm{K}) \,.
\end{equation}

$(\cdot)^{\top}$ denotes matrix transpose. Similarly, we can easily represent the total number of people who need to move in the system:
\begin{equation}
sum_{K} = \sum_{i} \bm{K} .
\end{equation}
The objective function is then given by the following equation, where a small positive constant $\varepsilon$ was used to avoid division by zero (set to be $1 \times 10^{-6}$). 
\begin{equation}
obj(\bm{K}) = \frac{sum_{AK}}{sum_{K} + \varepsilon}.
\end{equation}

Our optimization process is then represented as: 
\begin{equation}
\begin{aligned}
&\min_{\bm{K}}\; obj(\bm{K}),\\
& \text{subject to Constraints (1), (2), (3), (4)}.
\end{aligned}
\end{equation}

A suitable solver (e.g., an interior-point method) can be used to solve the above optimization problem and return the optimal matrix \(\bm{K}^{\star}\).

\section{Experiment}
Our optimization model was developed using Python 3.8.8, Numpy1.21.0, networkx 2.5, etc. for data preprocessing. The nonlinear optimization was developed using CasADi 3.6.7\footnote{https://web.casadi.org/}. The program was deployed on a Linux server (Architecture: x86\_64; CPU: 128 AMD Ryzen Threadripper
3990X 64-Core Processor).
\subsection{Study Area and Datasets}
We conducted the case study on the the railway network of Greater Tokyo, which is dominated by the world's most extensive urban rail network . Greater Tokyo area is made up of Tokyo and the three neighbor prefectures of Kanagawa, Saitama, and Chiba. This area is home to around 30\% of Japan's total population. It is the most populated metropolitan area in the world, with a population of 41 million in 2024 \footnote{https://en.wikipedia.org/wiki/Tokyo}. The network has a high rail transit usage ratio, with up to 48\% of the residents commute by train or metro services \cite{Jeph2022}. In this research, we collected central area of the railway network in which 90 lines, 21 operators of operational track and 1113 stations are included. We also used passenger count data for the year 2019 \cite{hangli2022multi,ge2024k}. The traffic data are the average count of passengers per day in one year\footnote{https://www.toukei.metro.tokyo.lg.jp/homepage/ENGLISH.htm}.


\subsection{Sub-graphs partition for Computational Acceleration}\label{sub-graph}

\begin{figure}[h]
    \centering
    \includegraphics[height=1.7in, width=0.75\linewidth]{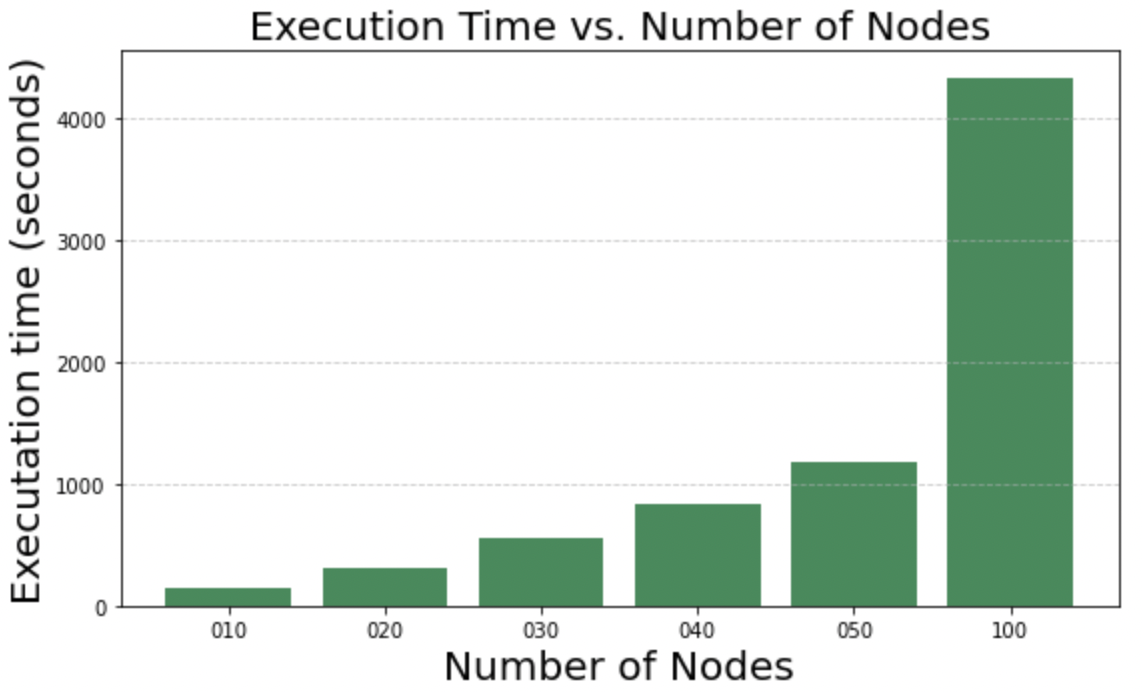} 
    \caption{The execution time increases significantly as the number of nodes increases.}
    \label{fig:exetime}
\end{figure}

Considering the large-scale graph size making them impractical, we proposed a sub-graphs partition solution for computational acceleration. Based on the cost matrix of $A_{cost}$, a $k$ nearest nodes sampling strategy was deployed to partition the entire graph into sub-graphs, where each sub-graph has $k$ nodes.  As a simple example, for each node $v_{i}$, the adjacency matrix is partitioned into small size matrix $\mathcal A_{i} \subseteq \mathbb{R}^{k\times k}$, where only consists of $k$ nodes. This $k$ nearest nodes strategy not only speeds up the computation but also corresponds well with the practical evacuation solution. Figure~\ref{fig:exetime} shows the comparison of computation time of optimization according different setting of K. Especially notable is the sharp rise when using 100 nodes, compared to much lower times for smaller node counts (e.g., under 500 seconds for 10–30 nodes). Hence, $k$ = 30 was chosen as the baseline for the following evaluations.

\subsection{Real World Check-in Data Preprocessing}
\begin{figure}[h]
\centering
\includegraphics[height=1.2in, width=0.8\linewidth]{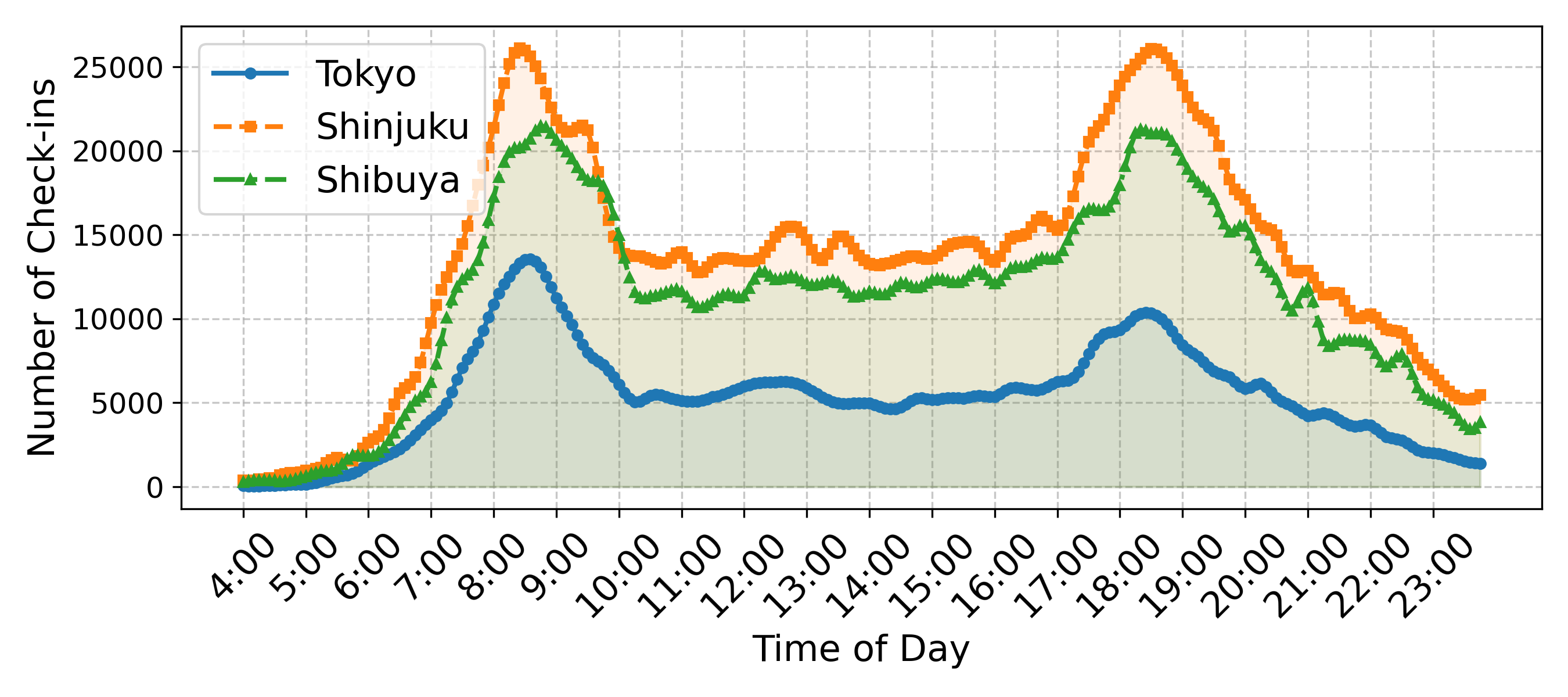}
\caption{Check-in counts dynamics at Tokyo, Shinjuku, and Shibuya Stations}
\label{fig:checkins}
\end{figure}

To obtain station check-in characteristics that accurately reflect the real-world model, we employed Blogwatcher footfall GPS data to sample the number of visitors within a 125m × 125m grid cell encompassing each station at 5-minute intervals. We then aggregated the data by averaging over three weekdays and adjusted it using the actual check-in data recorded at the station. Specifically, we derived a deflation coefficient by calculating the ratio of the total daily according to different time duration. We further applied the coefficient to the check-in pattern from the GPS data to generate dynamic station check-in data. Finally, constrained by the time range of the footfall data, we obtained station check-in records at 5-minute intervals from 4:00 AM to 11:45 PM. In Figure~\ref{fig:checkins}, we visualized the real world check in data of Tokyo, Shinjuku and Shibuya stations, which are ones of our main experiments targets.
We can observe two clear peaks (morning \& evening rush hours), which correspond to the commuting hours: people traveling to work/school in the morning and returning home in the evening. This figure effectively illustrates diurnal mobility patterns in urban transit, highlighting the dominance of rush-hour traffic and approximate to the real passenger counts of the stations.
\subsection{Experimental Setting}

\begin{figure}[h]
    \centering
    \includegraphics[height=2in, width=0.8\linewidth]{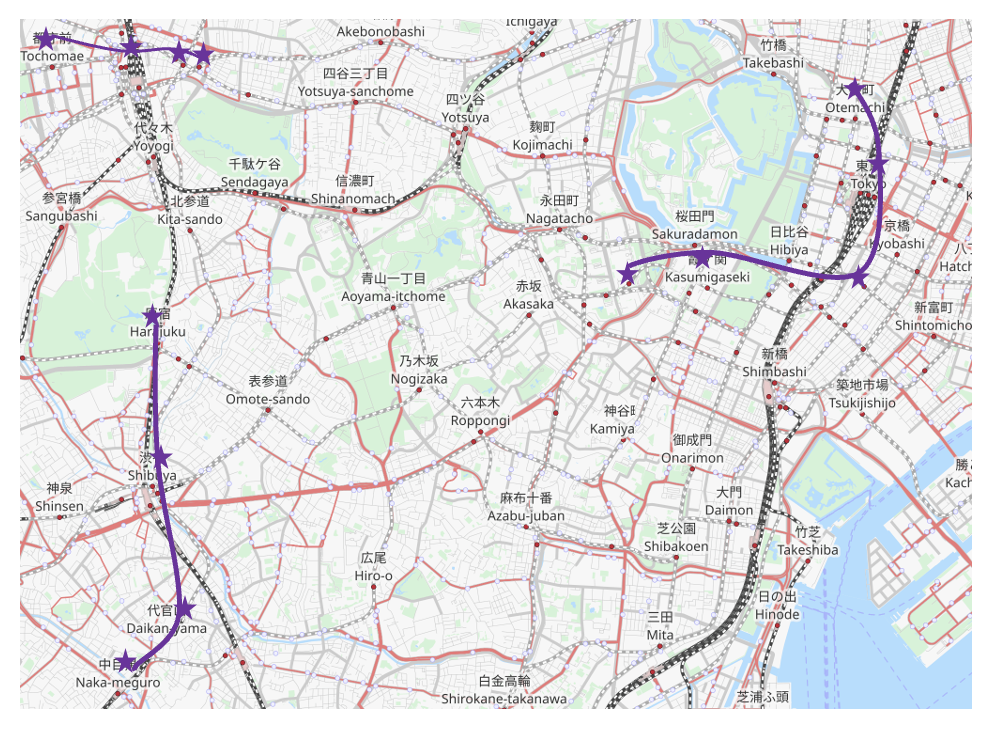} %
    \caption{The selected stations were highlighted as {\LARGE\color{violet} $\star$}; The map is based on the transportation view of OpenStreetMap, where the red dots show other stations}
    \label{osm}
\end{figure}

We conducted simulations on three sets of stations, each including several major stops: \{Otemachi, Tokyo, Ginza, Kasumigaseki, Kokai-gijidomae\}, \{Tochomae, Shinjuku, Shinjuku-sanchome, Shinjuku-gyoemmae\}, and \{Harajuku, Shibuya, Daikan-yama, Naka-meguro\} (Figure~\ref{osm}). In each simulation, we assumed that the selected line serving the corresponding station set was shut down for $T_{lm}$ (30 minutes), and all passengers from those stations needed to be evacuated to alternative stations. The affected passenger count was sampled by each 30 minutes based on average daily passenger count. Extracted from the statistic data, the effected passenger count within $T_{lm}$ of Tokyo, Shinjuku, and Shibuya are listed in Table~\ref{tab:checkin}. The maximum passenger volume capacity, which was set to be 1.5 in this experiment. We selected three time patterns, which is 8am, 12am, 6pm for validation.

\section{Evaluation}


\subsection{Evaluation Metrics}
We evaluate the simulation performances using three following metrics:
\begin{itemize}
    \item EPF (Evacuated Passenger Flow) is count of evacuated passengers which was optimized from the origin station to target stations.
    \item ATT (Average Travel (Cost) Time) is calculated as the total sum of cost time of origin station to destination station, to be divided by the total sum of evacuated passenger count.
\end{itemize}
\begin{table}[h]
\centering
\caption{Sampled Passenger Counts}
\begin{tabular}{p{3.0cm}lll}
\toprule
     & Tokyo & Shinjyuku & Shibuya \\\hline
Daily Average & 33488&  81639 &  67609 \\
Checkin Count of 8am  & 73271 & 145320&  115333  \\
Checkin Count of 12pm & 28534 & 79772 & 69687   \\
Checkin Count of 18pm & 46827&  135408&  110432 \\
\bottomrule
\end{tabular}
\label{tab:checkin}
\end{table}

\subsection{Results}
\begin{table}[h]
\centering
\caption{ATTs of different departure time}
\begin{tabular}{p{3.0cm}lll}
\toprule
     & Tokyo & Shinjyuku & Shibuya \\\hline
ATT (mins) of 8am  & 6.7   & 5.9       & 8.1     \\
ATT (mins) of 12am & 9.0   & 7.0       & 9.6     \\
ATT (mins) of 6pm & 5.2   & 6.0       & 5.1    \\
\bottomrule
\end{tabular}
\label{tab:results}
\end{table}

The ATT (Table~\ref{tab:results}) demonstrates the final optimized average travel time. The results are under 10 minutes, meaning it takes on average less than 10 minutes to evacuate passengers for evacuation. 
\begin{figure}[h]
    \centering
    \includegraphics[height=1.8in, width=0.9\linewidth]{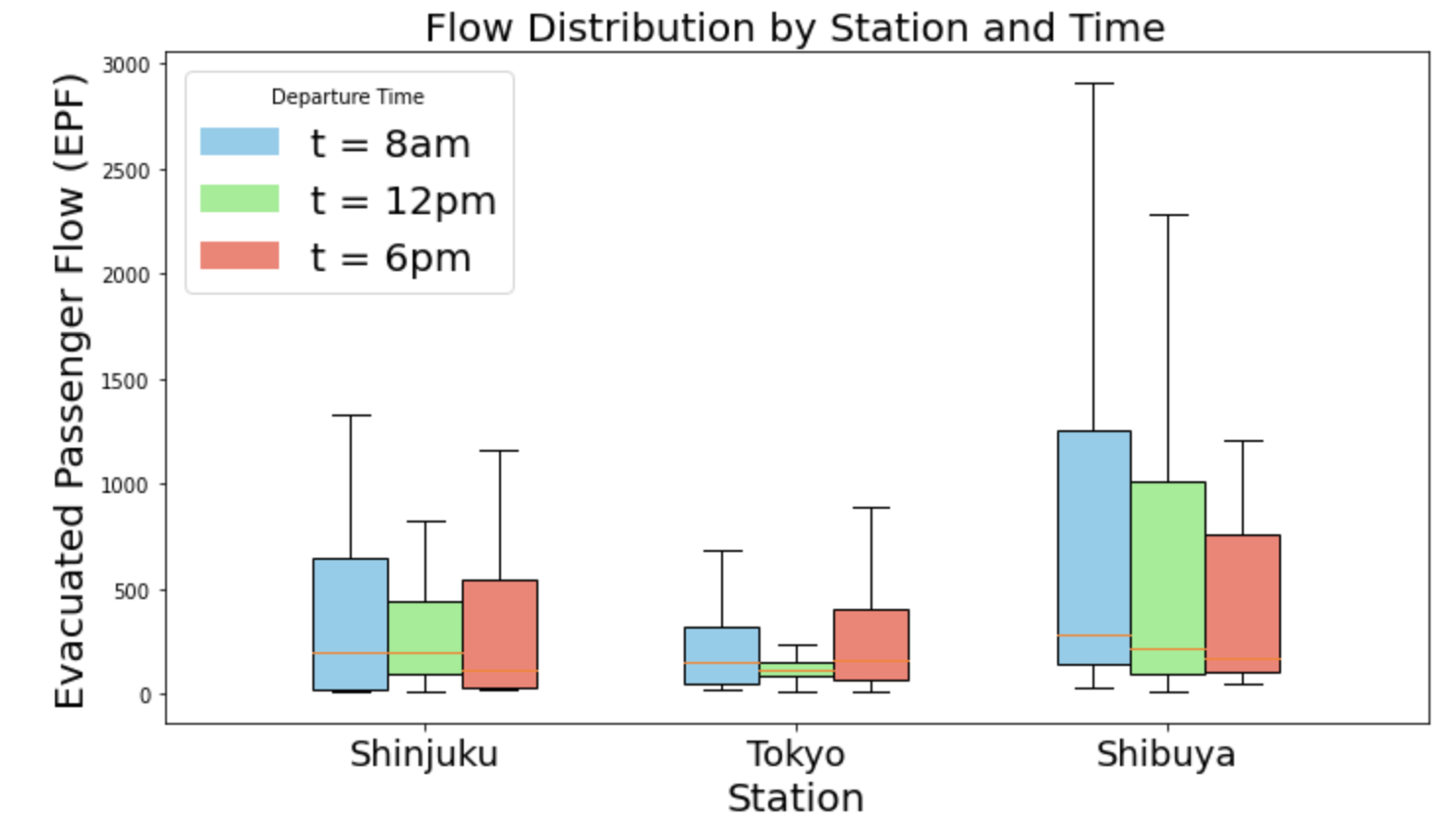} 
    \caption{Box plots of station-level evacuated passenger flow (EPF) at three experimental departure times.}
    \label{box}
\end{figure}
Aligning with Figure~\ref{box}, we can observe that comparing to Shibuya, Tokyo and Shinjuku stations are easier to evacuate passengers due to the presence of more nearby stations, whereas passengers at Shibuya station must travel longer time to reach other stations.

\begin{figure}[h]
    \centering
    \includegraphics[height=3.5in, width=0.9\linewidth]{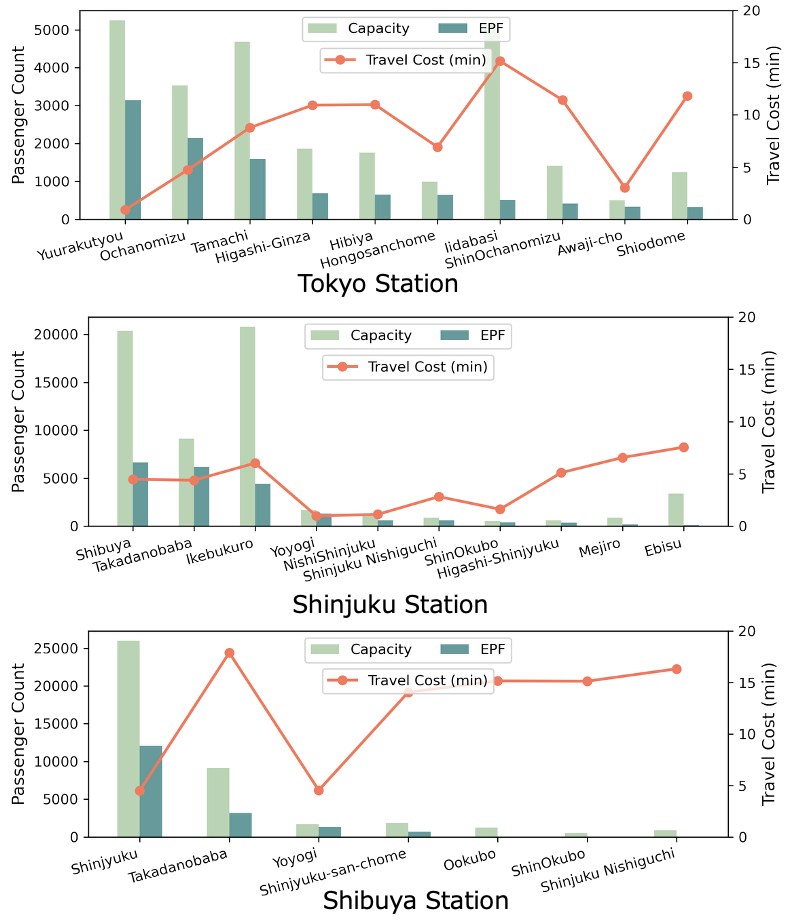} 
    \caption{Capacity and EPF of target stations from the origins of Tokyo, Shinjuku and Shibuya station, along with the corresponding travel cost.}
    \label{stations}
\end{figure}
 

Figure~\ref{stations} visualizes the capacity and evacuated passenger flow (EPF) of target stations. we can observe that directly linked stations are generally the preferred choices, as  passengers can travel to these stations in a very short time. Compared to Shinjuku and Shibuya, Tokyo station has more connectivities makes the majority of the evacuated target stations are the connected stations, followed by the geographically close stations. In the case of Shinjuku and Shibuya, some relatively large capacity but unconnected stations are chosen as evacuation target due to less connectivities.

\section{Conclusion}
In this paper, we have presented a traffic evacuation model for railway disruptions to improve resilience. Our proposed model focuses on optimizing evacuation based on the disruption of several lines of the rail network. We defined a multi-weighted geographic feature-fused travel cost matrix, which dynamically changes based on predefined interruption times. We proposed an optimization model using nonlinear programming, which considers the capacity of surrounding evacuation stations and the travel cost. In addition, our model can flexibly simulate disruptions at any number of stations and lines, enhancing its applicability.  We simulated our model utilizing the Tokyo's large-scale railway network ans tested three major stations. The result shown the evacuation time during emergencies is optimized, demonstrating the effectiveness of our approach. In the future work, applying reinforcement learning for fine-grained dynamic optimization could be considered.


\section*{Acknowledgment}
This work was supported by JSPS KAKENHI Grant Number JP25K21205. The work of Francesco Flammini was partly supported by the Swiss State Secretariat for Education, Research and Innovation (SERI) under contract no. 24.00528 (PhDs EU-Rail project). The project has been selected within the European Union’s Horizon Europe research and innovation programme under grant agreement no. 101175856. The views and opinions expressed in this paper are those of the authors only and do not necessarily reflect those of their affiliated institutions or the funding agencies.

\bibliography{IEEEabrv,ref}

\bibliographystyle{IEEEtran}

\end{document}